\documentclass[a4paper]{article}

\usepackage{19lomcon}        
\usepackage{cite}             
\usepackage{epsfig}           
\usepackage{epstopdf}
\usepackage[utf8]{inputenc}
\begin{document}


\title{Simulation of an experiment on looking for sterile neutrinos at nuclear reactor}
\author{S.~V.~Silaeva and V.~V.~Sinev
\vspace{3mm}
\email{vsinev@inr.ru}
}

\affiliation{
Institute for Nuclear Research of the Russian Academy of Sciences, Moscow 117312, Russia
\vspace{5mm}
}


\date{}
\maketitle


\begin{abstract}
The simulation of an experiment on looking for sterile neutrinos at a nuclear reactor at short distances is presented. It has been shown that statistical fluctuations in experimental bins always imitate the oscillatory behavior of the spectrum. An amplitude of the detectable oscillations decreases when statistics grows up in case of oscillations absence, while mass parameter tends to be accidental. When we simulate spectra in a detector with oscillations the parameters found in fitting become close to parameters applied to spectra starting from statistics 10$^5$ events in near detector.
\end{abstract}

\section{Introduction}
The neutrino oscillations hypothesis has been confirmed in a number of experiments with solar \cite{ahmad}, atmospheric \cite{abek} and reactors \cite{andb}$-$\cite{soob} neutrino. The parameters of all three transitions between active types of neutrino have been identified. However, in some experiments \cite{atha}$-$\cite{ment} abnormal results have been observed, which do not fit into the scheme of the three types of neutrino. The concept of the sterile neutrino has been introduced in order to describe these phenomena.

Various experiments to search for oscillations in the sterile state are being proposed \cite{gorb}$-$\cite{egor} at short distances, some of them have are underway already. In the other hand an analysis on sterile neutrinos is doing in many ongoing experiments.

A simulation of nuclear reactor antineutrino spectrum registration in a virtual experiment with two detectors looking for sterile neutrinos is presented here. We have chosen popular distances to reactor core (10-15 m) and assume there are two detectors installed at 12 and 14 m to the reactor core. Reactor core is accounted as point like. The goal was to obtain the value of statistics that allows to get oscillation parameters with 5$\sigma$ confidence level.

\section{Simulation of the observed spectrum of reactor antineutrino}

Antineutrino interacts with the hydrogen nuclei by means of the reaction of inverse beta-decay (IBD) in a detector:

\begin{equation}
\bar{\nu_{e}} + p \rightarrow n + e^{+}
\end{equation}

The reaction’s energy threshold (1) is 1.806 MeV. The positron energy is linked with the energy of antineutrino as follows:
\begin{equation}
E_{\nu} = T_{e} + E_{thr} + r_n
\end{equation}
where $E_{\nu} -$ antineutrino energy, $T_{e} -$ positron kinetic energy, $E_{thr} -$ reaction threshold (1) and $r_n -$ neutron recoil energy. The neutron energy is small and in the first approximation can be neglected.
 
In the presence of oscillations, the positrons spectrum in the detector should be changed in accordance with the theory of neutrino oscillations. Oscillations to sterile mode should be seen at distances much smaller than for oscillations of active neutrino flavours. Minimal distance for oscillation of $\nu_{e} \rightarrow \nu_{\tau}$ is hundreds meters. But sterile neutrinos are looking for at tens meters to the reactor centre. In this case we can use a formulae for oscillations between only two neutrino states proposed by B. Pontecorvo not taking into account other flavour oscillations.  

The probability to conserve its flavor for electron antineutrino with energy $E_{\nu}$ at a distance $L$ from the reactor core can be written as:
\begin{equation}
P(\nu_e \rightarrow \nu_e)= 1 - \sin^22\theta \cdot \sin^2\left(1.267\cdot \frac{\Delta m^2 [\textrm{eV}^2]\cdot L[\textrm{m}]}{E_{\nu} [\textrm{MeV}]}\right), 
\end{equation}
where $\sin^22\theta -$ oscillation amplitude, 1.267 $-$ the coefficient reconciling units in the expression of sine function, $L -$ the distance to the source of antineutrinos, $\Delta m^2 -$ mass squared difference for oscillating neutrino states, $E_{\nu} -$ antineutrino energy.

If there are no oscillations, then the shape of antineutrino spectrum (as well as positron one) is not changed with a distance from neutrino source and integral of the spectrum satisfies to $1/L^2$ law. In the presence of oscillations at specific distances depending on neutrino energy according to (3), the spectrum’s shape is modified and the law of inverse distance squared is violated. On very large distances from the source (more than several dozens of oscillation lengths (4)), the spectrum becomes the same as the source again and its integral reduced by factor $\frac{1}{2}\sin^22\theta$ The oscillations’ length is determined by value of mass parameter and neutrino’s energy:
\begin{equation}
L_{osc}= \frac{E_{\nu}}{1.267 \cdot \Delta m^2 }
\end{equation}

\begin{figure}[t]
\centering
\includegraphics[height=7cm]{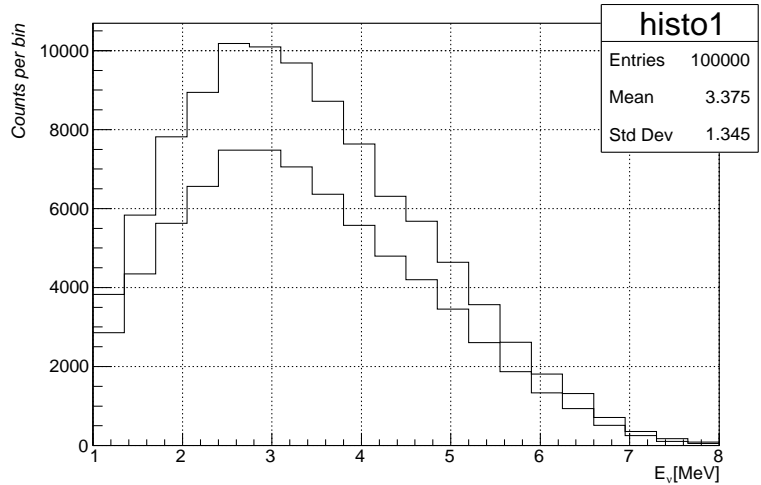}
\caption{Positron spectra for two distances. Larger spectrum for distance of $L_1 =$ 12 m, lower one $–$ for distance of $L_2 =$ 14 m.}
\end{figure}

For simulations we used the antineutrino spectrum from \cite{sine}. Antineutrino spectrum was transferred into the positron one multiplying by cross section of the IBD reaction (1). For the cross section we used simplified formula from the \cite{stru}
\begin{equation}
\sigma_{\nu p}= 10^{-43} p_e E_e E^{-0.07056+0.02018\cdot\textrm{ln}E_{\nu}-0.001953\cdot\textrm{ln}^3 E_{\nu}}_{\nu}, E_e = E_{\nu} - 1.2933, 
\end{equation}
where $p_e$ and $E_e$ are momentum and total energy of the positron from inverse beta decay reaction, $E_{\nu} -$ antineutrino energy.

The positron spectrum was simulated with with chosen statistics $N_1$ at near distance (12 m) and $N_2 =\left(\frac{L_1}{L_2}\right)^2$ for far position (14 m). In this case we have a pair of spectra at two distances and can analyse the influence of oscillations. We regard these spectra as experimental ones. Spectra were simulated as histograms with energy step 0.31 MeV.

For the oscillation analysis we made ratio of “experimental” spectra and fit it with theoretical ratio of oscillation functions.

\begin{equation}
R_i = \frac{N_{i}(L_2)}{N_{i}(L_1)}, 
\end{equation}
where $N_{i}$ is content of $i$-th bin for corresponding spectrum.

The ratio of oscillations probability for chosen distances is written as a separate function, which used for fitting the simulated (“experimental”) spectra: 
\begin{equation}
f_{12}(E_{\nu_e}) = \frac{P(L_2)}{P(L_1)}=\frac{1 - \sin^22\theta \cdot \sin^2\left(1.267\cdot\frac{\Delta m^2 \cdot L_2}{E_{\nu}}\right)}{1 - \sin^22\theta \cdot \sin^2\left(1.267\cdot\frac{\Delta m^2 \cdot L_1}{E_{\nu} }\right)} 
\end{equation}

The analysis was made by minimization of $\chi^{2}$ function constructed as follows:
\begin{equation}
\chi^{2}= {\sum}_i{\frac{(R_i-f_{12,i}(\Delta m^2,\sin^22\theta))^2}{{\sigma}^2_{1i}}} +\frac{(\frac{I_2}{I_1}-\frac{P_2}{P_1}\cdot (\frac{L_1}{L_2})^2)^2}{{\sigma}^2_{2i}}, 
\end{equation}
where $R_i$ $-$ a ratio of positron spectra for two distances determined at (6), $f_{12,i}(\Delta m^2,\sin^22\theta)$ $-$ function of oscillation probabilities ratio (7), ${\sigma}_{1i}$ $-$ uncertainty of bin's ratios (6), ${\sigma}_{2i}$ $-$ experimental uncertainty of simulated spectra integrals ratio, $I_1$ and $I_2$ $-$ integrals of simulated spectra, and $P_1$ and $P_2$ $-$ integrals of oscillation functions.

\section{Virtual experiment on looking for sterile neutrinos oscillations}

\subsection{Simulation of spectra without oscillations}

To understand what oscillations could be observed in an experiment with two detectors placed close to nuclear reactor the pairs of spectra were simulated. The distances 12 and 14 m between detector and reactor core centers were chosen. The simulated spectra for statistics 10$^5$ events of the spectrum in the near detector and statistics corrected on a distance for far position are shown in figure 1. 

The analysis on looking for oscillations was done by fitting the ratio of histograms by the expression (7). Spectra ratio for statistics 10$^5$ events in each spectrum for distances 12 m and 14 m is shown in figure 2. The fitted oscillation curve is also shown here.

The simulation of ten spectrum pairs has been done for the event statistics 10$^5$ in the near detector. Founded oscillation parameters after minimisation of $\chi^{2}$ function (8) are shown in table 1. As one can see from table 1 values of mass parameter $\Delta m^2$ are grouped around 1 eV$^2$ in range 0.4$-$1.5 eV$^2$, and the parameter sin$^{2}2\theta$ is approximately equal to the statistical error. 

\begin{figure}[t]
\centering
\includegraphics[height=7cm]{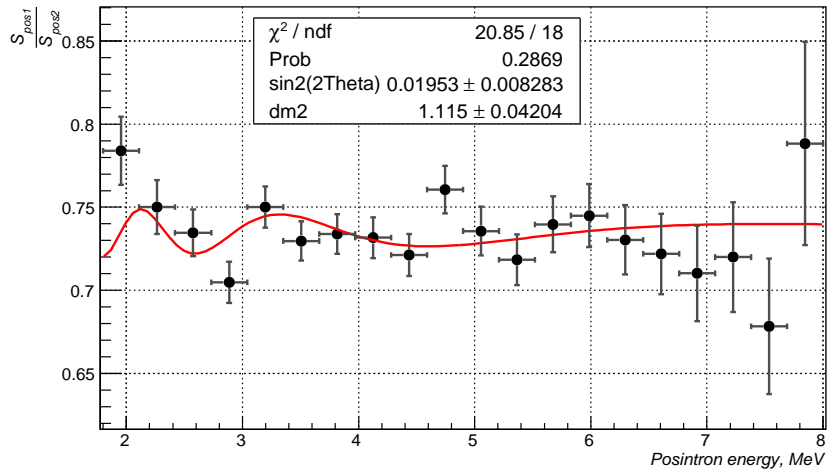}
\caption{Simulated positron spectra ratio without oscillations. Red line – the result of functions ratio fitting (6). The oscillation parameters were founded as a result of value minimization, are shown in the figure insertion.}
\end{figure}

\begin{table}[ht]
\caption{Oscillation parameters found from the analysis of simulated spectra couples based on statistics 10$^5$ events for the closest detector position.}
\centering
\vspace{2mm}
 \begin{tabular}{| c | c | c |} 
 \hline
 $\Delta m^2$, eV$^2$ & sin$^{2}2\theta$ & $\chi^{2}_{min}$ \\
 \hline
 0.685 & 0.036 & 19.4 \\ 
 \hline
 0.885 & 0.005 & 9.7 \\ 
 \hline
 0.972 & 0.006 & 12.5 \\ 
 \hline
 1.442 & 0.018 & 8.2 \\ 
 \hline
 1.266 & 0.030 & 9.5 \\ 
 \hline
 0.943 & 0.027 & 8.7 \\ 
 \hline
 0.447 & 0.042 & 19.9 \\ 
 \hline
 1.478 & 0.011 & 13.0 \\ 
 \hline
 0.655 & 0.013 & 8.1 \\ 
 \hline
 0.940 & 0.007 & 23.0 \\ 
 \hline
\end{tabular}
\end{table}

\begin{table}[ht]
\caption{Values of the oscillation parameters $\Delta m^2$ and sin$^{2}2\theta$ at varied statistics in simulated spectra at different statistics without oscillations.}
\centering
\vspace{2mm}
 \begin{tabular}{| c | c | c | c |} 
 \hline
 Events number & $\Delta m^2$, eV$^2$ & sin$^{2}2\theta$ & $\chi^{2}_{min}$ \\
 \hline
 10$^4$ & 0.770 & 0.956 & 3.9 \\ 
 \hline
 10$^5$ & 1.266 & 0.030 & 9.5 \\ 
 \hline
 10$^6$ & 1.247 & 0.005 & 21.3 \\ 
 \hline
\end{tabular}
\end{table}

\begin{table}[ht]
\caption{Deviation from the minimum value of $\chi^{2}_{min}$ for the Gaussian distribution with two degrees of freedom.}
\centering
\vspace{2mm}
 \begin{tabular}{| c | c | c |} 
 \hline
 $N\sigma$ & ${\Delta}{\chi}^2$ & $P$,\% \\
 \hline
 1.0 & 2.30 & 68.27 \\ 
 \hline
 1.64 & 4.61 & 90.0 \\ 
 \hline
 2.0 & 6.18 & 95.45 \\ 
 \hline
 3.0 & 11.83 & 99.73 \\ 
 \hline
 4.0 & 19.35 & 99.9937 \\ 
 \hline
 5.0 & 28.44 & 99.999943 \\ 
 \hline
\end{tabular}
\end{table}

\begin{figure}[t]
\centering
\includegraphics[height=10cm]{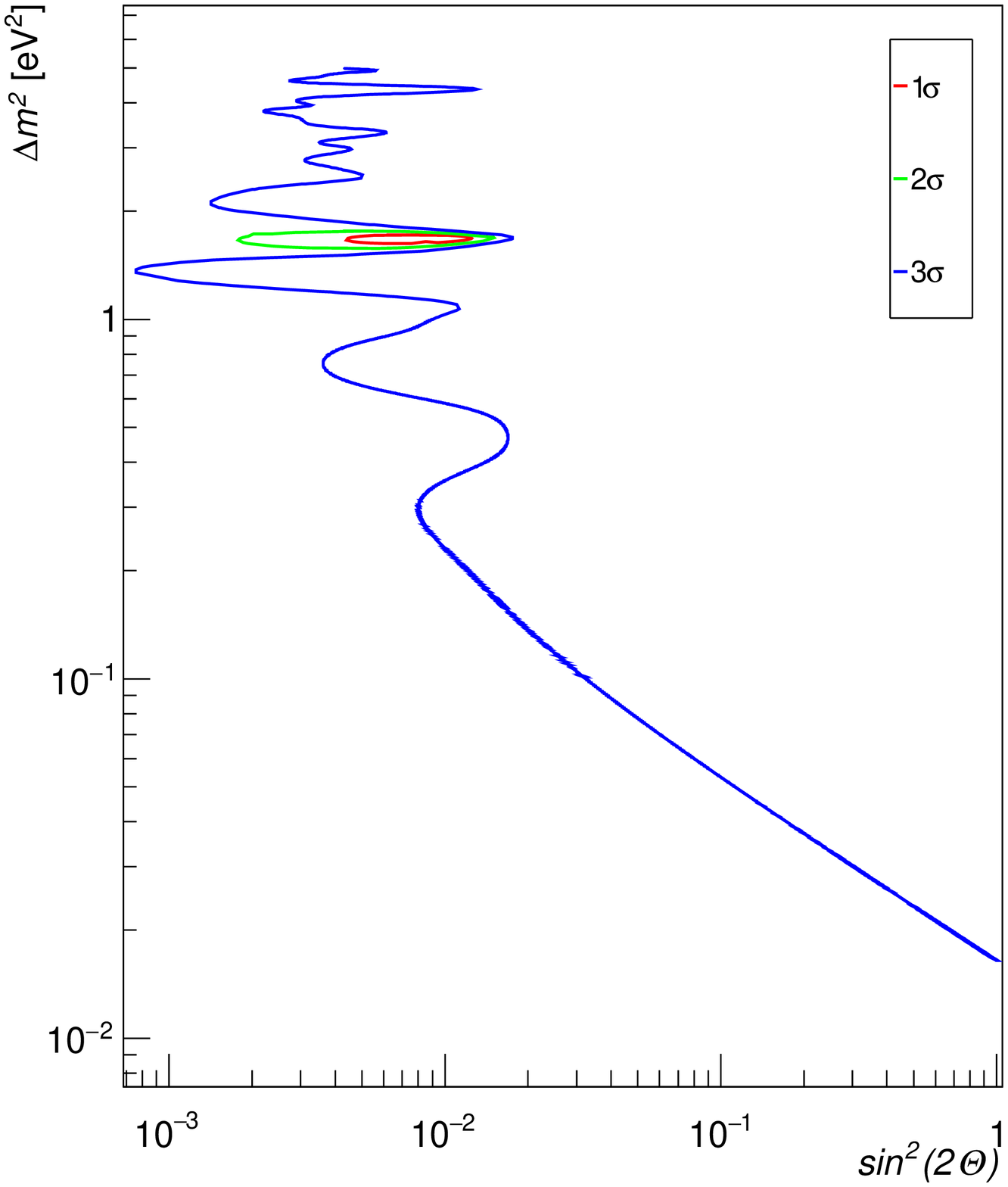}
\caption{Oscillation parameters boundaries for statistics 10$^4$ events within spectrum for the closest distance. The shown areas: red line $-$ 1$\sigma$, green line $-$ 2$\sigma$ and blue line $-$ 3$\sigma$.}
\end{figure}

The influence of statistics on  value of oscillation parameters was studied. Pairs of spectra were simulated for the event statistics 10$^4$, 10$^5$ and 10$^6$ events. The results are shown in table 2. We see that with statistics increasing the parameter sin$^{2}2\theta$ goes down in case of oscillations absence. The mass parameter $\Delta m^2$ gets accidental values but close to 1 eV$^2$.

The oscillation parameters boundaries are shown on figure 3 for statistics 10$^6$ events. It was found that statistics fluctuations in bins always form false oscillations. In case of oscillations absence the parameters may be found at level of 2$-$2.5 standard deviations. The curves of limitations on oscillation parameters were made by looking for parameters that correspond to difference $\Delta\chi^2 = \Delta\chi^2_{min} + n$, where values of $n$ were taken from table 3 for Gaussian distribution for two degrees of freedom.

\subsection{Simulation with oscillations}

Verification of necessary statistics for detecting oscillations was a goal for the second phase of the simulation. 

Mass parameter $\Delta m^2 = 2$ eV$^2$ was found in \cite{ment}. In multiple reactor neutrino experiments at 80th$-$90th years of last centure mass parameter found in oscillation analysis varied in a region 0.5$-1.5$ eV$^2$.

We decided to make simulations with parameters close to reactor neutrino experiments. The spectra in a detector were simulated assuming possible oscillations with parameters $\Delta m^2 = 1$ eV$^2$ and sin$^{2}2\theta = 0.05$.

The analysis was done according to the method described above. We have repeated simulations for two detectors placed at the same distances as in simulations without oscillations. Ten pairs of spectra were simulated for the same distances as without oscillations for the events’ statistics 10$^5$ events in the near detector. The results of the analysis are shown in the tables 4 and 5. We can see that parameters of oscillations that were applied certainly found here.    

\begin{table}[ht]
\caption{Oscillation parameters found from the analysis of simulated pairs of spectra based on statistics 10$^5$ events for the near detector position with oscillations $\Delta m^2 = 1$ eV$^2$ and sin$^{2}2\theta = 0.05$.}
\centering
\vspace{2mm}
 \begin{tabular}{| c | c | c |} 
 \hline
 $\Delta m^2$, eV$^2$ & sin$^{2}2\theta$ & $\chi^{2}_{min}$ \\
 \hline
 1.045 & 0.059 & 18.2 \\ 
 \hline
 1.004 & 0.058 & 10.9 \\ 
 \hline
 0.994 & 0.049 & 12.0 \\ 
 \hline
 0.991 & 0.046 & 10.2 \\ 
 \hline
 1.011 & 0.045 & 10.7 \\ 
 \hline
 0.997 & 0.069 & 9.4 \\ 
 \hline
 0.977 & 0.061 & 17.9 \\ 
 \hline
 1.050 & 0.043 & 10.5 \\ 
 \hline
 0.982 & 0.048 & 8.0 \\ 
 \hline
 1.005 & 0.051 & 20.9 \\ 
 \hline
\end{tabular}
\end{table}

\begin{table}[ht]
\caption{Values of the oscillation parameters in simulated spectra at different statistics with used parameters $\Delta m^2 = 1$ eV$^2$ and sin$^{2}2\theta = 0.05$.}
\centering
\vspace{2mm}
 \begin{tabular}{| c | c | c | c |} 
 \hline
 Events number & $\Delta m^2$, eV$^2$ & sin$^{2}2\theta$ & $\chi^{2}_{min}$ \\
 \hline
 10$^4$ & 0.879 & 0.070 & 5.7 \\ 
 \hline
 10$^5$ & 1.011 & 0.045 & 10.7 \\ 
 \hline
 10$^6$ & 1.036 & 0.048 & 20.6 \\ 
 \hline
\end{tabular}
\end{table}

The ratio of simulated spectra including oscillations in the antineutrino energy scale is shown at figure 4 (statistics 10$^5$ events). The resulting curves for oscillation parameters limitations are show at figure 5 for statistics 10$^6$ events. We see that parameter regions continue to be enclosed up to 5$\sigma$.

\begin{figure}[t]
\centering
\includegraphics[height=7cm]{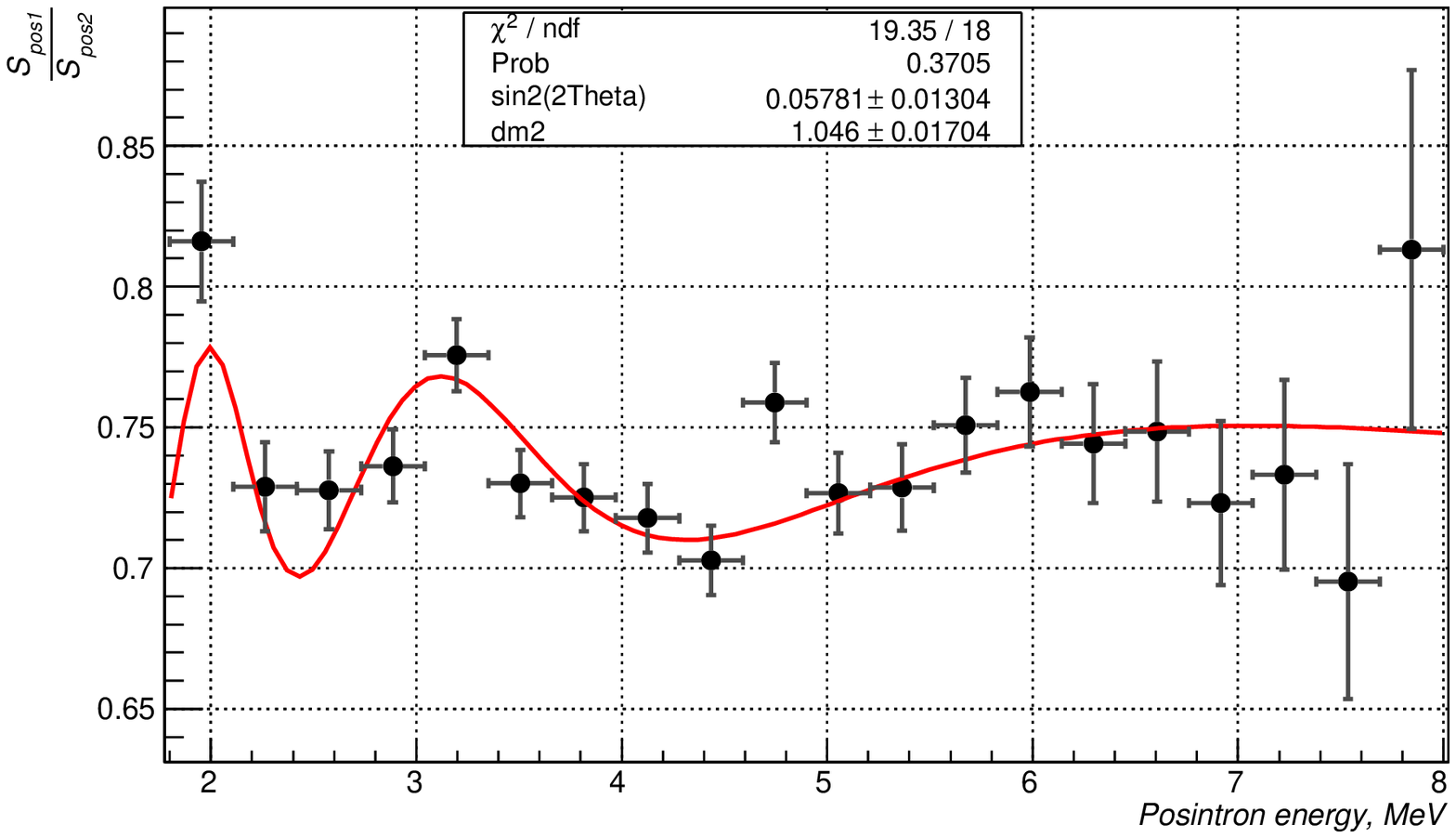}
\caption{Simulated positron spectra ratio with chosen oscillations. Red line – the result of functions ratio fitting. The oscillation parameters were founded as a result of value minimization, are shown in the figure insertion.}
\end{figure}

\begin{figure}[t]
\centering
\includegraphics[height=10cm]{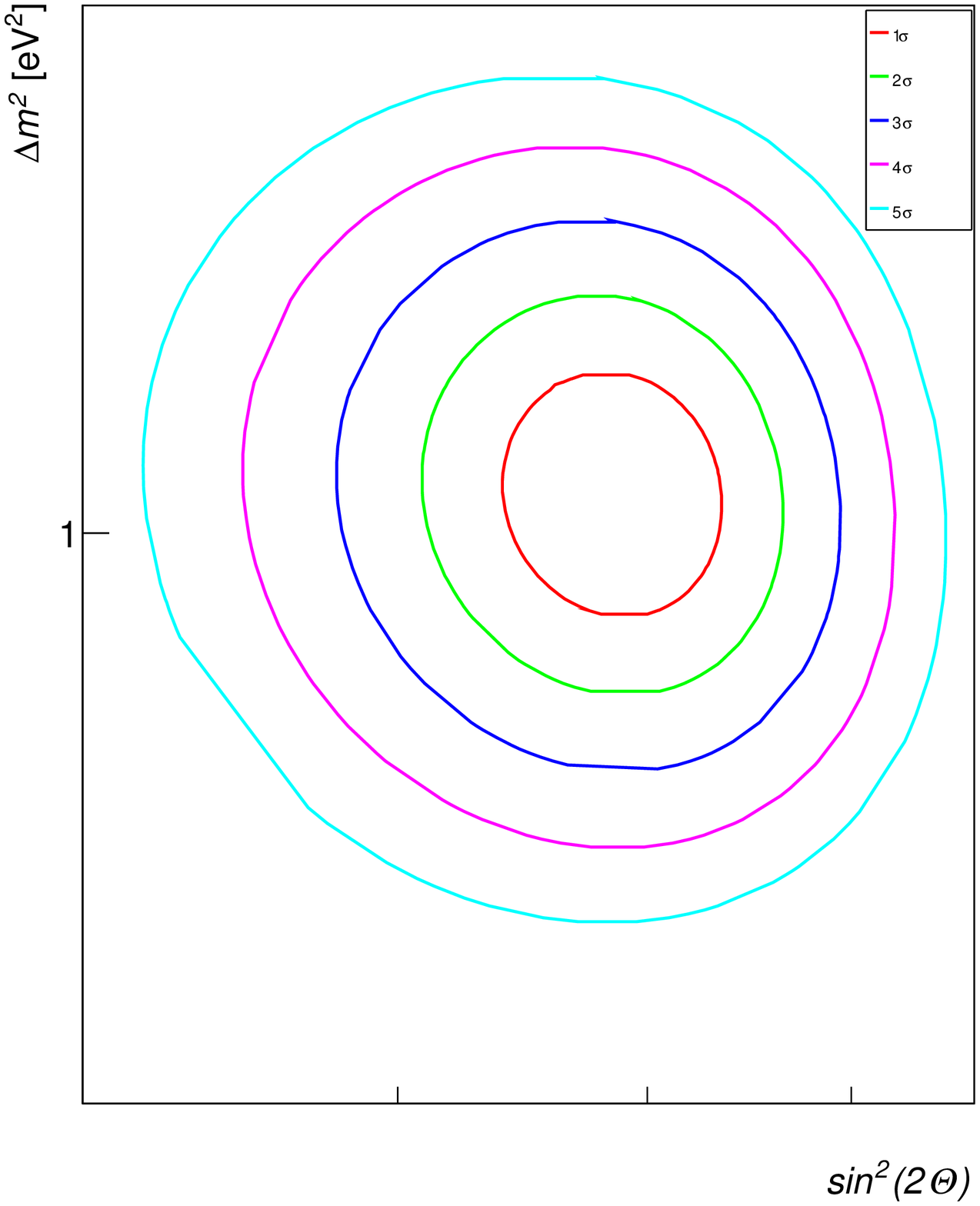}
\caption{Oscillation parameters boundaries for statistics 10$^6$ events for the spectrum at near position. The shown areas: red line $-$ 1$\sigma$, green line $-$ 2$\sigma$ and blue line $-$ 3$\sigma$. Oscillation parameters are $\Delta m^2 = 1$ eV$^2$ and sin$^{2}2\theta = 0.05$.}
\end{figure}

\section{Conclusion}

The simulation of an experiment on searching for sterile neutrinos at nuclear reactor was done. The virtual experiment with two detectors placed at 12 and 14 m is presented. The pairs of positron spectra with definite statistics with and without oscillations were simulated. Oscillations were accepted with the parameters of $\Delta m^2 = 1$ eV$^2$ and sin$^{2}2\theta = 0.05$. In current work chosen distances are close to the distances used in the DANSS experiment \cite{egor}. 

Simulated spectra were analysed as experimental ones. Ratio of spectra at two distances was fitted with ratio of oscillation functions. 

The analysis of simulated spectra with oscillations has shown that at minimal statistics ($\sim 10^4$ of events in each spectrum) the oscillations could be detected with 95\% C.L. The statistics do not allow to determine applied oscillations on the higher confidence level. At the events’ statistics 10$^5$ events the confidence level in four standard deviations can be reached, but it is not come to 5$\sigma$. However, it is possible to reach 5$\sigma$ confidence level in an experiment with two detectors having statistics above 10$^6$ neutrino events at each distance.

In case of oscillations absence at any statistics one can find oscillation parameters areas at the level of $2-2.5\sigma$. Moreover, founded oscillation parameters have demonstrated the following behavior: when statistics is growing the parameter sin$^{2}2\theta$ is decreased to achieving the value corresponding to statistical uncertainty, and the parameter $\Delta m^2$ has shown random values in the range from 0.1 up to 10 eV$^2$.

At this stage, the systematic error was not taken into consideration. 

In real experiment the systematic error should be added to the statistical one thus the statistics should be increased in order to identify the oscillations effect on the confidence level of 5$\sigma$.

\section*{Acknowledgments}
Authors are grateful to L. B. Bezrukov for useful discussions.


\end{document}